\documentclass[12pt]{article}

\usepackage{graphicx} 
\usepackage{jheppub}
\usepackage{amsfonts,amsmath} 
\usepackage{mathrsfs} 
\usepackage{setspace}
\usepackage[utf8]{inputenc} 
\usepackage{comment} 

\makeatletter
\def\@fpheader{\relax}
\makeatother

\DeclareMathAlphabet{\mathbbold}{U}{bbold}{m}{n} 
\newcommand{\be}{\begin{equation}} \newcommand{\ee}{\end{equation}}

\newcommand{\thalf}{{\tfrac{1}{2}}}

\DeclareMathOperator{\Tr}{Tr}

\title{On the Kerr-AdS/CFT correspondence}

\author[a,b]{Juli\'{a}n Barrag\'{a}n Amado,}\emailAdd{j.j.barragan.amado@rug.nl}
\author[a]{Bruno Carneiro da Cunha,}\emailAdd{bcunha@df.ufpe.br}
\author[b]{and Elisabetta Pallante}\emailAdd{epallante@rug.nl}

\affiliation[a]{Departamento de F\'{i}sica, Universidade Federal de Pernambuco,
50670-901, Recife, Pernambuco, Brazil} 
\affiliation[b]{Van Swinderen Institute for Particle Physics and
  Gravity, University of Groningen, Groningen, The Nederlands} 

\abstract{We review the relation between four-dimensional global
  conformal blocks and field propagation in ${\rm AdS_5}$. Following
  the standard argument that marginal perturbations should backreact
  in the geometry, we turn to the study of scalar fields in the
  generic Kerr-${\rm AdS_5}$ geometry. On one hand, the result for
  scattering coefficients can be obtained exactly using the
  isomonodromy technique, giving exact expressions in terms of $c=1$
  chiral conformal blocks. On the other hand, one can use the analogy
  between the scalar field equations to the Level 2 null field Ward
  identity in two dimensional Liouville field theory to write
  approximate expressions for the same coefficients in terms of
  semi-classical chiral Liouville conformal blocks. Surprisingly, the
  conformal block thus constructed has a well-behaved interpretation
  in terms of Liouville vertex operators.}


\begin{document}

\maketitle

\section{Introduction}

The gauge/gravity correspondence has opened up a new avenue to study
non-perturbative phenomena in field theories at conformal fixed points
and some patterns of conformal symmetry breaking, using
(semi)-classical gravitational analogues. In 20 years since its
inception, a plethora of different applications has sprung, from exact
results pertaining to integrable sectors in supersymmetric Yang-Mills
(SYM) to phenomenological model building. In most of those, the role
of black hole backgrounds as an analogue of a thermal state is
paramount\,---\,see \cite{marolf2009black} for a review. In fact,
perturbations of the thermal state, or analogously perturbations of
the black hole, are fundamental to the understanding of the
lowest-lying spectrum of the underlying theory.  

In many of those models, the perturbation comes in the guise of a
semi-classical scalar field in a classical black hole
background. It serves as a mock-up model of true interactions in
large-N ${\cal N}=4$ SYM and some deformations of it
\cite{Girardello:1998pd,DeWolfe:1999cp}, like ${\cal 
  N}=2$ SYM with generic matter multiplets in the adjoint and/or
fundamental. In general relativity, it also tests linear stability of
solutions \cite{Berti:2009kk}, relaxation times
\cite{Horowitz:1999jd}, as well as relation with the Conformal Field
Theory (CFT) bootstrap, represented by Liouville field theory 
\cite{Jackson:2014nla}.  Most of the applications, however, either
focus on low dimension, with the Bañados-Teitelboim-Zanelli (BTZ)
black hole \cite{Banados:1992wn} serving as a ubiquitous background,
or on higher-dimensional solutions without rotation\,---\,a serious
drawback for many interesting physical phenomena, like superradiance,
the zero temperature moduli and the fate of the inner horizon. 

One powerful method to study rotating black holes in any dimension
is the monodromy method, introduced in this context by
\cite{Neitzke:2003mz, Motl:2003cd}. The method has deep ties to
integrable systems and the Riemann-Hilbert problem in complex
analysis, relating scattering coefficients to monodromies of a flat
holomorphic connection associated to the radial differential equation
obtained after separation of variables. Recently, one of the authors
helped uncover the relation between scattering coefficients of some
Kerr-de Sitter \cite{Novaes:2014lha,daCunha:2015uua} black holes in
four dimensions to exact $c=1$ conformal blocks using deformations of
the flat holomorphic connection\,---\,isomonodromy deformations. These
conformal blocks were in itself computed exactly
\cite{Gamayun:2013auu} in terms of the Painlevé VI $\tau$-function,
capitalizing on recent developments in gauge/gravity correspondence in
the form of the Alday-Gaiotto-Tachikawa (AGT) conjecture
\cite{Alday:2009aq,Alba:2010qc}.  

The relation between scattering amplitudes of black holes in any
dimension and two-dimensional conformal blocks has been alluded to by
many authors and used to obtain bulk information from the boundary
dynamics, see \cite{Heemskerk:2012mn,Castro2013,Castro2013b,
  Hijano:2015zsa, Fitzpatrick:2015zha}. The purpose of the present
work is to review the relation of semi-classical (global) conformal
blocks in four dimensions to propagating (scalar) fields in
five-dimensional anti-de Sitter (AdS), a fact outlined by
\cite{daCunha:2016crm} and \cite{Hijano:2015zsa,Czech:2016xec,deBoer:2016pqk}
revisiting work by Ferrara and collaborators in the 70's
\cite{Ferrara:1971vh,Ferrara:1972xq,Ferrara:1973vz}. The
construction does not rely on supersymmetry. We will also be  
interested in the exact relations between two-dimensional conformal
blocks and scattering of scalar fields in a Kerr-AdS background in
five dimensions, whose greybody factors are also given in terms of the
Painlevé VI $\tau$-function. Due to this mathematical fact, the
background displays a hidden conformal symmetry\,---\,albeit
not directly connected to the one presented in
\cite{Castro:2010fd}. In an appropriately defined semiclassical limit,
these blocks are described by Liouville field theory and the
corresponding vertex operators are matched.  

The paper is structured as follows. In Section \ref{4dCFT}, we review
the work of \cite{daCunha:2016crm} and give the ${\rm AdS_5}$
interpretation of four-dimensional CFT correlation functions\,---\,see
also \cite{Ferrara:1973vz}\,---\,and close by discussing the
modifications allowed by asymptotically ${\rm AdS_5}$ spaces. In
Subection \ref{5dKerr}, we turn to the five-dimensional Kerr-AdS black
hole, and write the scalar perturbation equation of motion in terms of
two Heun ordinary differential equations (Fuchsian differential
equations with 4 regular singular points). In Subsection
\ref{isomonodromy}, we present formal solutions for the black hole normal
modes in terms of the Painlevé VI $\tau$-function, thus relating them to
$c=1$ exact conformal blocks. In Section \ref{liouville}, we point out that
the near extremal limit should be well described by semi-classical (global) 
conformal blocks, which in turn are well described by semi-classical
Liouville field theory. We use this analogy to construct an
equivalence between the solutions for the scalar field in Kerr-${\rm
  AdS_5}$ to Liouville 4-point correlators, with insertions associated
to the singular points of the differential equation. We point out
that, while the singular points of the angular equation, as well as
the insertions associated to the point at infinity and the
``unphysical'' point $r_0$ in the radial equation are associated to
usual Liouville vertex operators, the insertions associated to the
inner and outer horizons are on the Seiberg bound line. We close by
discussing the usefulness and relevance of the results for model
building and outline the prospect of future work.

\section{CFTs and Scalar Fields in ${\rm AdS}_5$}
\label{4dCFT}

\begin{figure}[htb]
\begin{center}
\includegraphics[width=0.8\textwidth]{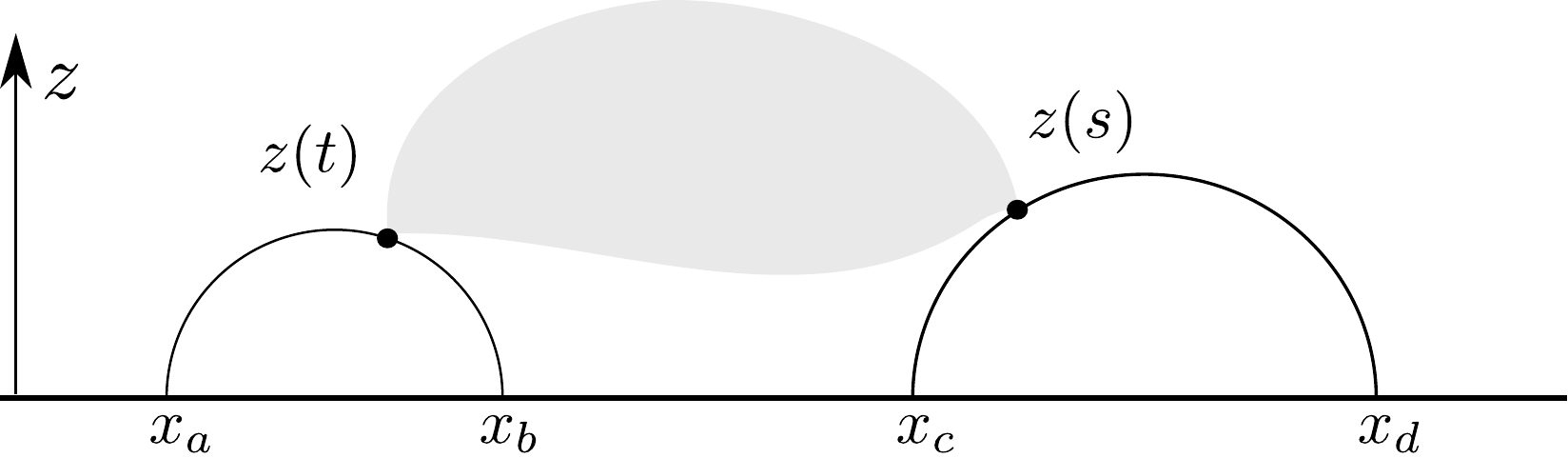}
\caption{The boundary classical conformal block in the AdS
  representation. While the points at spatial infinity ($z=0$) are
  linked by geodesics (the semicircles), the points at the geodesics
  are linked by the scalar field propagator. The mass $\mu$ of the
  propagator is related to the intermediate channel conformal weight
  $\Delta$ by \eqref{eq:mutodelta}.}
\label{fig:witten4pt}
\end{center}
\end{figure}

We review a few well-known facts about scalar fields in AdS space.
The reader unfamiliar with the conformal bootstrap in generic
dimensions may find the review in \cite{SimmonsDuffin:2012uy} useful. 
We will begin the discussion by considering the conformal block of a
generic dual conformal field theory, 
defined by the projection of the four point function of primary
(spinless) operators onto the conformal descendants of (the equally
spinless) ${\cal O}$,
\begin{equation}
Q(x_a,x_b,x_c,x_d)=\langle {\cal O}_a(x_a) {\cal O}_b(x_b) P_{\cal O}{\cal
  O}_c(x_c) {\cal O}_d(x_d) \rangle .
\label{eq:ccb}
\end{equation}
It has been verified that, under generic conditions
\cite{Ferrara:1972xq,Hijano:2015zsa,deBoer:2016pqk,daCunha:2016crm}, 
$Q$ is given by the expression:
\begin{multline}
Q=K\,C_{ab\Delta}C_{\Delta cd}
((x_{ab})^2)^{-\tfrac{\Delta_a+\Delta_b}{2}}((x_{cd})^2)^{-\tfrac{\Delta_c+\Delta_d}{2}}
  \\ 
\times
\int_0^\infty\frac{dt}{t^{\tfrac{\Delta_a-\Delta_b}{2}+1}} 
\int_0^\infty\frac{ds}{s^{\tfrac{\Delta_c-\Delta_d}{2}+1}} 
\langle \Phi_\Delta (z(t), x(t)) \Phi_\Delta(z(s),x(s))\rangle_{0},
\label{eq:witten4pt}
\end{multline}
where $t=(1-u)/u$ and $s=(1-v)/v$, with $u$ and $v$ spanning the
geodesics as above, see also Fig. \ref{fig:witten4pt}. The constant
$K$ is a product of beta functions:
\begin{equation}
K=\frac{4^{d-\Delta}\Gamma(\Delta)^2}{\Gamma(\thalf(a\Delta b))
\Gamma(\thalf(b\Delta a)) \Gamma(\thalf(c\Delta d))
\Gamma(\thalf(d\Delta c))}.
\end{equation}
The variables $\Phi_\Delta$ are constructed from the intermediate
primary fields ${\cal O}_\Delta$ from a boundary-to-bulk procedure --
again we refer to the literature cited above for details. The function
between brackets
\begin{equation}
\langle \Phi_\Delta(z_1,x_1)\Phi_\Delta(z_2,x_2)\rangle_0 =
G_n(z_1,x_1;z_2,x_2)
\end{equation}
denotes a particular integral transformation of the two-point function
of ${\cal O}_\Delta$ in a conformally invariant ``vacuum-state'' in
the CFT. 

The expression for $Q$ has a natural interpretation as the propagation
of a scalar field in a ${\rm AdS}_{d+1}$ background, with metric given
by
\begin{equation}
ds^2=\frac{dz^2+\eta_{\mu\nu}dx^\mu dx^\nu}{z^2}.
\end{equation}
where the mass is related to the conformal dimension of the
intermediate channel by
\begin{equation}
\Delta=d/2+\sqrt{d^2/4+\mu^2}.
\label{eq:mutodelta}
\end{equation}
In the Poincaré metric, the semicircles in Fig. \ref{fig:witten4pt}
are geodesics, and the bulk-to-bulk propagator between the internal
points is given by the free scalar field in ${\rm AdS}_{d+1}$:
\begin{equation}
G_n(\sigma)=2^{-\Delta}\sigma^{-\Delta}{_2F_1}(\thalf\Delta,\thalf\Delta+\thalf
;\Delta -\tfrac{d-2}{2};\sigma^{-2}),
\end{equation} 
with
\begin{equation}
\sigma=2z-1
=\frac{z_1^2+z_2^2+\eta_{\mu\nu}(x_1-x_2)^\mu(x_1-x_2)^\nu}{2z_1z_2}. 
\label{eq:poincaresigma}
\end{equation}
related to the geodesic distance by $\sigma=\cosh(\ell)$. We define
$z$ for future reference.

The results above establish a geometric interpretation of conformal
systems in any dimension, and open up the possibility of studying
conformal symmetry breaking mechanisms through the dynamics of
geometry, a tool widely used in general relativity. This point of view
is supported via the Maldacena
conjecture, where some (relevant) perturbations of the CFT 
vacuum state $|0\rangle$ can be modelled through the deformation of
the background AdS space-time. Formulas like the boundary-to-bulk
propagator as well as the Witten diagram
\eqref{eq:witten4pt} can be suitably modified by changing the path
of integration and the bulk-bulk propagator, in order to accommodate for the
change of geometry. If the space-time is only asymptotically anti-de
Sitter, then the exact formulas above will only hold when the
separations between points are small, {\it i.e.}, in the UV
limit. Asymptotically AdS spaces will then correspond to UV conformal
fixed points in the boundary theory.  

Even beyond SUSY and exact backgrounds, there is some evidence to
trust that quantum corrections can be geometrized in a way similar to
above, at least for a class of conformal field theories and for some
patterns of conformal symmetry breaking. In particular, this should be
true for boundary field theories where the decoupling from the
lowest-lying low-energy spectrum of the theory happens as described in 
\cite{Fitzpatrick:2013twa}. In this case, for instance, if we take the
``cluster-decomposition'' limit of \eqref{eq:witten4pt} where 
\begin{equation}
x_a\simeq x_b,\quad x_c\simeq x_d,\quad |x_c-x_a|\gg 1,
\end{equation}
we can think of the ${\rm AdS}$ propagator as giving the would-be
meson-meson potential. Another important question is if
zero-temperature QCD realizes such decoupling. The results put forward
in \cite{Bochicchio:2014vna} for large-N Yang-Mills say that the
logarithmic corrections to conformal symmetry induced by asymptotic
freedom in the scalar glueball two-point function differ from the UV
behaviour of AdS/CFT realizations. At the same time, confinement and
asymptotic freedom are two aspects of one single phenomenon of
conformal symmetry breaking.  The latter may be illustrated by a naive
geometric ``hard wall'' construction, for which the discrete would-be
meson mass spectrum is given by
\begin{equation}
\frac{p_n^2}{\Lambda^2}=-x_{\nu,n},
\end{equation}
where $\Lambda = 1/z_0$ is an infrared cutoff and $x_{\nu,n}$ is the
n-th zero of the regular Bessel function $J_\nu(x)$,  has nevertheless
phenomenologically appealing features and could still serve as a
description of the deconfined phase of QCD, in the presence of mass
(temperature) perturbations. In other words, the successes in
describing quark-gluon plasma resonances via AdS/CFT 
genuinely  depend on the deconfined and near-conformal nature of the
high-temperature system, where the conformally-invariant Coulomb
potential is corrected by temperature and mass perturbations.  

Certainly, the geometrization described in this Section points to
the fact that there is a class of broken-CFTs that are described by an
asymptotically AdS space-time. Admittedly, the conditions presented here
are much weaker than having, for instance, a generic background where
quantum corrections are under control. Such class of broken-CFTs
should at least include the deconfined region of (supersymmetric)
non-Abelian gauge theories described in the classical tests of
AdS/CFT, as well as many examples of low-dimensional condensed matter
systems. 

In the next section, we consider rotating black hole solutions in AdS
as a tool to describe thermal field theories, and to deviate from pure
AdS towards the infrared. The relevant quantities on the geometric
side of the correspondence are basically restricted to the black hole
mass, angular momenta and whichever monopole charges we include in the 
theory. By analyzing this restricted and still tractable space of
parameters, one could hope to learn more about a broad spectrum of
dual CFTs.

\subsection{The Kerr-${\rm AdS}_5$ background}
\label{5dKerr}

We pick the simplest asymptotically AdS background in five dimensions,
Kerr-${\rm AdS}_5$. From the argument of the last section, its
purported dual field theory will display a (trivial) conformal fixed  
point in the UV, with no mass gap. Scalar field perturbations around
this background will give the would-be scalar meson spectrum of the
boundary theory, as per the correspondence built in the preceding
section, eventually allowing us to describe a class of breaking
patterns of the boundary CFT in the UV.  On the other hand, the IR
behavior will be radically different from pure AdS, with the horizon
now playing the role of a thermal background. One notes two new
features of the analysis presented here. Firstly, the holographic
interpretation of the space-time charges is not unique, it will depend
on the conformal structure we choose for the asymptotic metric. The
choice consists, from a four dimensional perspective, to consider the
CFT living either on a spatial plane or a sphere. Analyzing the three elements of
the Cartan subalgebra of ${\rm SO(4,2)}$, we see that in both choices
the time-translation is kept, so the energy is the same in either of
them. The angular momenta, though, can be thought of as values from
translational momenta or one component of the translational momentum,
say $p_z$, and the angular momentum parallel to it. We will take the
latter interpretation. Secondly, there is a non-trivial moduli of
parameters even at zero temperature: in five dimensions there is a
one-parameter family of zero temperature black holes which can be
thought to represent non-trivial conformal fixed points in the IR of
the dual theory. 

Also, some thermodynamical comments are in order. Because of the
rotation, the Killing vector associated with the outer horizon becomes
space-like sufficiently far from the source, so that any thermal
radiation would have to move faster than light to maintain equilibrium.
The analysis of \cite{Hawking:1998kw} has shown that thermodynamical
equilibrium is possible for Kerr-AdS black holes in a region of the
parameter space, and there exist phase transition(s) to a region where
the boundary manifold rotates faster than the speed of light
at least in one direction.  A duality is also inferred
\cite{Hawking:1998kw} between the Kerr-AdS black hole at critical
angular velocities and the CFT (in one dimension less) at the boundary
of AdS space. The extremal Kerr-${\rm AdS}_5$ black hole has been
further thermodynamically studied in \cite{Lu:2008jk}, revealing
another class of holographic duality. The equality of thermodynamical
quantities (entropy) suggests a duality of the extremal Kerr-AdS black
hole (in any dimension) and a chiral 2d CFT for each rotation (see
also \cite{Guica:2008mu}).  Thus, for rotating and asymptotically AdS
black holes in five dimensions, one expects two types of holographic
duality: one with a four-dimensional CFT on the rotating manifold at
the boundary of AdS, using AdS/CFT correspondence,  the other with
chiral two-dimensional CFTs (one for each rotation) in the extremal
black hole limit, using Kerr/CFT correspondence in the near horizon
region. As we will see, the analysis presented here will point to a
third option.

The Kerr-${\rm AdS}_5$ metric is \cite{Hawking:1998kw}
\begin{align}
ds^{2} &=
         -\dfrac{\Delta_{r}}{\rho^{2}}\left(dt-\dfrac{a_1\sin^{2}\theta}{1-a_1^2}d\phi
         -\dfrac{a_2\cos^{2}\theta}{1-a_2^2}d\psi\right)^{2}+\dfrac{\Delta_{\theta}
         \sin^{2}\theta}{\rho^{2}}\left(a_1\,dt-\dfrac{(r^{2}+a_1^{2})}{1-a_1^2}d\phi
         \right)^{2} \nonumber  \\ 
&+ \dfrac{1+r^{2}}{r^{2}\rho^{2}}\left( a_1a_2\,dt -
  \dfrac{b(r^{2}+a_1^{2})\sin^{2}\theta}{1-a_1^2}d\phi -
  \dfrac{a(r^{2}+a_2^{2})\cos^{2}\theta}{1-a_2^2}d\psi \right)^{2}
  \nonumber \\ 
&+ \dfrac{\Delta_{\theta}\cos^{2}\theta}{\rho^{2}}\left( a_2\,dt-
  \dfrac{(r^{2}+a_2^{2})}{1-a_2^2}d\psi \right)^{2}
  +\dfrac{\rho^{2}}{\Delta_{r}}dr^{2}+
  \dfrac{\rho^{2}}{\Delta_{\theta}}d\theta^{2}, \label{eq1}   
\end{align}
where
\begin{gather}
\Delta_{r} =
             \dfrac{1}{r^{2}}(r^{2}+a_1^{2})(r^{2}+a_2^{2})(1+r^2)-2M,
             \nonumber \\ 
\Delta_{\theta} = 1- a_1^2\cos^{2}\theta -
                  a_2^{2}\sin^{2}\theta, \quad\quad \ 
\rho^{2} = r^{2} + a_1^{2}\cos^{2}\theta + a_2^{2}\sin^{2}\theta,
\label{eq2}
\end{gather}
and $a_1$ and $a_2$ are two independent rotation parameters. This metric
satisfies $R_{\mu\nu} = -4g_{\mu\nu}$, and asymptotically
approaches AdS space with radius of curvature $\ell_{AdS}=1$. The mass
relative to the ${\rm AdS}_5$ vacuum solution and the angular momenta
are
\begin{equation}
{\cal M}=\frac{3\pi M}{4(1-a_1^2)(1-a_2^2)},\quad\quad
{\cal J}_\phi=\frac{\pi M a_1}{2(1-a_1^2)^2},\quad\quad
{\cal J}_\psi=\frac{\pi M a_2}{2(1-a_2^2)^2}.
\end{equation}

The horizons of the black hole are obtained from the equation
$\Delta_{r} = 0$, which can be reduced to a cubic equation by making
the substitution $x=r^{2}$. For real $a_1$, $a_2$ and positive $M$ there are two
real roots to this equation, $r_{+}^{2}$ and $r_{-}^{2}$. The largest
of these roots, $r_{+}^{2}>r_{-}^{2}$, corresponds to the outer event
horizon \cite{Gibbons:2004ai}. 
Note that this metric possesses three commuting Killing vectors,
\begin{equation}\label{eq4}
\xi_{k} = \dfrac{\partial}{\partial t} +
\Omega_{a}(r_{k})\dfrac{\partial}{\partial \phi} +
\Omega_{b}(r_{k})\dfrac{\partial}{\partial \psi} 
\end{equation}
such that they are null at each respective horizon and has the time
translational and rotational (bi-azimuthal) isometries. This entails
to the constants $\Omega_{k} \equiv \Omega(r_{k})$ being the angular
velocities of each horizon and one can interpret these quantities as
angular velocities in two independent orthogonal 2-planes of
rotation. The temperature and angular velocities of each horizon for
an observer following $\xi_{k}$ orbits, with $k=+,-,0$, are given
by 
\begin{gather}
\Omega_{k,1} = \dfrac{a_1 (1-a_1^2)}{r^{2}_{k} + a_1^{2}} \qquad
  \Omega_{k,2} = \dfrac{a_2 (1-a_2^2)}{r^{2}_{k} + a_2^{2}} \nonumber \\ 
T_{k} = \dfrac{r^{2}_{k}\Delta'(r_{k})}{4\pi(r^{2}_{k} +
  a_1^{2})(r^{2}_{k} + a_2^{2})}
= \frac{r_k}{2\pi}\frac{(r_k^2-r_i^2)(r_k^2-r_j^2)}{(r_k^2+a_1^2)(r_k^2+a_2^2)}.
\label{eq5} 
\end{gather}

We note for future reference that for reasonable values of $M$, $a_1$
and $a_2$, $r_+$ and $r_-$ are real and positive, while
$r_0^2=-r^2_--r_+^2-a_1^2-a_2^2-1$ is negative. By the same token, $T_+$
is positive, $T_-$ is negative and $T_0$ is purely imaginary. In the
extremal case $r_+=r_-$, $T_\pm$ goes to zero linearly with
$r_+^2-r_-^2$ while $T_0$ is finite. We also note that \eqref{eq1}
reduces to the global five-dimensional AdS metric when $M=a_1=a_2=0$ and
that the metric displays a singularity hidden by an event horizon when
$T_+\ge T_-\ge 0$. 

\subsection{Kerr-anti de Sitter Wave Equation}
One of the most interesting properties of Kerr-(A)dS black hole
(\ref{eq1}), with arbitrary rotation parameters in five dimensions, is
the separability of the Hamilton-Jacobi equation for a particle, and
also the separability of the Klein-Gordon (KG) equation for a
scalar field in this background
\cite{Kunduri:2005fq}. We consider now the KG equation with mass $\mu$
together with the Ansatz $\Phi =
\Pi(r)\Theta(\theta)e^{-i\omega t + im_{1}\phi +im_{2}\psi}$ where $m_{1}$,
$m_{2}$ $\in \mathbbold{Z}$.  This mode solution results in
two decoupled ordinary differential equations for radial and angular 
functions. The angular equation is given by 
\begin{multline}
\dfrac{1}{\sin\theta\cos\theta}\dfrac{d}{d\theta}\left(\sin\theta
  \cos\theta\Delta_{\theta}  
  \dfrac{d\Theta(\theta)}{d\theta}\right) 
-\left[\omega^2+\frac{(1-a_1^2)m_1^2}{\sin^2\theta}+
  \frac{(1-a_2^2)m_2^2}{\cos^2\theta} \right. \\ \left.
  -\frac{(1-a_1^2)(1-a_2^2)}{\Delta_\theta}(\omega+m_1a_1+m_2a_2)^2 
+\mu^2(a_1^2\cos^2\theta+a_2^2\sin^2\theta)\right]\Theta(\theta)
=-\lambda\Theta(\theta),
\label{eq8}
\end{multline}
where $\lambda$ is the separation constant. For more details, see
\cite{Aliev:2008yk}, although we will present the necessary
ingredients to bring the angular equation to the canonical form of a
Heun equation.  Thus, we use the transformation $u = \sin^{2}\theta$,
under which the four regular singular points are located at 
\begin{equation}
u=0,\quad u=1,\quad u=u_{0}=\frac{1-a_1^2}{a_2^2-a_1^2},\quad u=\infty, 
\end{equation}
and the critical exponents are defined as the asymptotic behavior of
the function near the singular points $S(u)\simeq (u-u_i)^{\alpha_i}$
or $S(u)\simeq u^{-\alpha_\infty}$ for the point at infinity
\begin{gather}
\alpha_0=\pm\frac{m_1}{2},\quad \alpha_1=\pm\frac{m_2}{2},\quad
\alpha_{u_0}=\pm\frac{1}{2} (\omega+a_1m_1+a_2m_2),\quad
\alpha_\infty=1\pm\sqrt{1+\frac{\mu^2}{4}}.
\end{gather}
The sign choice is immaterial and we will take the positive solution
as standard. Let us now define $\beta=\omega+a_1m_1+a_2m_2$. Performing
the change of variables
\begin{equation}
\Theta(u) = u^{-m_1/2}(u-1)^{-m_2/2}(u-u_{0})^{-\beta/2}S(u)
\label{eq:shomos}
\end{equation}
we bring the angular equation to the canonical Heun form
\begin{equation}
\frac{d^2S}{du^2}+\left(\frac{1-m_1}{u}+\frac{1-m_2}{u-1}
  +\frac{1-\beta}{u-u_0} \right)\frac{dS}{du}+\left(
\frac{q_1q_2}{u(u-1)}+\frac{u_0(u_0-1)Q_0}{u(u-1)(u-u_0)}\right)S=0
\label{eq:heunangular}
\end{equation}
with the accessory parameters given by
\begin{equation}
q_1q_2=\tfrac{1}{4}(\beta+m_1+m_2-2)^2-1-\tfrac{1}{4}\mu^2
\label{eq:angularthetainfty}
\end{equation}
\begin{multline}
4u_0(u_0-1)Q_0=\frac{\lambda-\omega^2-\mu^2}{a_2^2-a_1^2} 
+(u_0-1)((m_1+\beta-1)^2-m_2^2-1)\\
+u_0((m_2+\beta-1)^2-m_1^2-1).
\label{eq:angularaccessory}
\end{multline}
One notes that \eqref{eq:heunangular} has the same AdS spheroidal
harmonics form as the problem in four dimensions.

The radial equation can be written in the form
\begin{multline}\label{eq9}
\dfrac{1}{r\Pi(r)}\dfrac{d}{dr}\left(r\Delta_{r}
  \dfrac{d\Pi(r)}{dr}\right) - \biggl[ \lambda + \mu^{2}r^{2} +
\dfrac{1}{r^{2}}(a_1a_2\omega - a_2(1-a_1^2)m_{1} -
a_1(1-a_2^2)m_{2})^{2}\biggr] \\
+ \dfrac{(r^{2}+a_1^{2})^{2}(r^{2}+a_2^{2})^{2}}{r^{4}\Delta_{r}}\left(
  \omega - \dfrac{m_{1}a_1(1-a_1^2)}{r^{2}+a_1^{2}} -
  \dfrac{m_{2}a_2(1-a_2^2)}{r^{2}+a_2^{2}}\right)^{2}  = 0. 
\end{multline}
This equation again has four regular singular points, located at the
roots of $r^2\Delta_r$ and infinity
\begin{equation}
r^2=r_-^2,\quad r^2=r_+^2,\quad r^2=r_0^2,\quad r^2=\infty,
\end{equation}
with critical exponents given in terms of the temperatures and angular
velocities by
\begin{equation}\label{eqThetas}
\theta_{k} = \pm \dfrac{i}{2\pi}\left(\dfrac{\omega -
    m_{1}\Omega_{k,1} - m_{2}\Omega_{k,2}}{T_{k}}\right), \quad
\theta_{\infty}=\frac{\Delta}{2},\frac{d-\Delta}{2} =1\pm
\sqrt{1+\frac{\mu^2}{4}}, 
\end{equation}
where $k=0,+,-$. As with the angular equation, the choice of root is
immaterial and we will choose the positive sign to define the
$\theta_k$. Note that $\theta_\infty$ is the scaling dimension
from the CFT interpretation as in the analysis of last section. To
bring this equation to the canonical Heun form, we perform the change
of variables 
\begin{equation}
z=\frac{r^2-r_-^2}{r_+^2-r_-^2},
\end{equation}
and write it in terms of 
\begin{equation}
\Pi(z) = z^{-\theta_{-}/2}(z-1)^{-\theta_{+}/2}(z-z_0)^{-\theta_{0}/2}R(z), 
\label{eq:shomor}
\end{equation}
where $z_0=(r_0^2-r_-^2)/(r_+^2-r_-^2)$. After some uneventful
algebraic manipulations, we arrive at the equation for $R(z)$
\begin{equation}
\dfrac{d^{2}R}{dz^{2}} + \biggl[\dfrac{1-\theta_{-}}{z}
+\dfrac{1-\theta_{+}}{z-1} +\dfrac{1 -\theta_{0}}{z-z_0} \biggr]\dfrac{dR}{dz} + 
\left( \frac{
    \kappa_0\kappa_1}{z(z-1)}+\frac{z_0(z_0-1)K_0}{z(z-1)(z-z_0)}\right)R(z)
=  0,
\label{eq:heunradial}
\end{equation}
where $z_0$ is given as above, and
\begin{equation}
\kappa_0\kappa_1=\frac{1}{4}(\theta_-+\theta_++\theta_0-2)^2
-1-\frac{\mu^2}{4}
\label{eq:radialthetainfty}
\end{equation}
\begin{multline}
4z_0(z_0-1)K_0=-\frac{\lambda+\mu^2r_0^2-\omega^2}{r_+^2-r_-^2}
+(z_0-1)[(\theta_0+\theta_--1)^2-\theta_+^2-1]\\
+z_0[(\theta_0+\theta_+-1)^2-\theta_-^2-1]
\label{eq:radialaccessory}
\end{multline} 

Both equations \eqref{eq:heunangular} and \eqref{eq:heunradial} can be
solved by usual Frobenius methods in terms of Heun series. We are,
however, interested in solutions which satisfy
\begin{equation}
S(u)=
\begin{cases}
1+{\cal O}(u),\quad\quad u\rightarrow 0, \\
1+{\cal O}(u-1),\quad\quad u\rightarrow 1,
\end{cases}
\label{eq:boundaryfors}
\end{equation}
which will set a quantization condition for the separation constant
$\lambda$. Note that $u_0$ in \eqref{eq:heunangular} is always greater
than one for the $a_2<1$ case, below the extremal limit. The condition that
$R(z)$ corresponds to a purely ingoing wave at the outer horizon $z=1$
and normalizable at infinity is
\begin{equation}
R(z)=
\begin{cases}
1+{\cal O}(z-1),\quad\quad z\rightarrow 1, \\
z^{-\Delta/2-(\theta_++\theta_-+\theta_0)/2}(1+{\cal
  O}(z^{-1})),\quad\quad z\rightarrow \infty, 
\end{cases}
\label{eq:boundaryforr}
\end{equation}
with $\Delta=2+\sqrt{4+\mu^2}$. This condition will enforce the
quantization of the (not necessarily real) frequencies $\omega$, which
will correspond to the (quasi)-normal modes as functions of the
conformal weight $\Delta$ and of the parameters $T_k$, $\Omega_{k,1}$
and $\Omega_{k,2}$. The rather difficult problem is that the
quantization condition cannot be expressed in terms of elementary
functions of the parameters of the equations \eqref{eq:heunangular}
and \eqref{eq:heunradial}. We turn to this problem now.

\subsection{Isomonodromy and formal solutions}
\label{isomonodromy}

Finding the quantization conditions from the problem above can be cast
in terms of a reverse Riemann-Hilbert problem. Details can be found in
\cite{Novaes:2014lha,daCunha:2015uua}, and we will present here a
summary of the ideas. Take a generic Fuchsian second order ordinary
differential equation (ODE),
\begin{equation}
y''(z)+p(z)y'(z)+q(z)y(z)=0.
\end{equation}
Frobenius method allows us to compute the solutions at
a particular singular point $z_k$
\begin{equation}
y^+_{k}(z)=(z-z_k)^{\alpha^+_k}(1+{\cal O}(z-z_k)),\quad\quad
y^-_{k}(z)=(z-z_k)^{\alpha^-_k}(1+{\cal O}(z-z_k)),
\label{eq:frobeniussolutions}
\end{equation}
where $\alpha^\pm_k$ are the solutions of the indicial equation at
$z=z_k$. We will call $y^\pm_{k}(z)$ constructed thusly the ``Frobenius
solutions'' at $z_k$. These Frobenius solutions $y_k^\pm(z)$ have the
property that, under analytical continuation around $z_k$, yield
\begin{equation}
y^{\pm}_{k}(e^{2\pi i}(z-z_k)+z_k)=e^{2\pi i\alpha^\pm_k}y^{\pm}_k(z).
\label{eq:frobeniusmonodromy}
\end{equation}

For our purposes, it will be more interesting to recast the Fuchsian
ODE in terms of the matrix differential equation, called ``Garnier system'',
\begin{equation}
\frac{d\Phi}{dz}=A(z)\Phi,
\end{equation}
where $\Phi(z)$ is a matrix of fundamental solutions
\begin{equation}
\Phi(z)=\begin{pmatrix}
y^{(1)}(z) & y^{(2)}(z) \\
w^{(1)}(z) & w^{(2)}(z)
\end{pmatrix},
\label{eq:14}
\end{equation}
where $y^{(1,2)}$ are any two linearly independent solutions of the
ODE\,---\,they can be our Frobenius solutions $y^\pm_k(z)$ as above, but
we will treat with the generic case\,---\,and $w^{(1,2)}$ are auxiliary
functions which depend explictly on the choice for the matrix $A(z)$
\begin{equation}
w^{(1)}(z)=\frac{1}{a_{12}(z)}\left(\frac{dy^{(1)}}{dz}-a_{11}(z)y^{(1)}\right),
\label{eq:wfromy}
\end{equation}
with analogous expressions for $w^{(2)}$. It is straightforward to see
that, if the solutions $y^{(1)}(z)$ and $y^{(2)}(z)$ are linearly
independent, then $\Phi(z)$ is invertible. The matrix 
\begin{equation}
A(z)=\frac{d\Phi}{dz}\Phi(z)^{-1}
\label{eq:flatconnection}
\end{equation}
can now then be thought of as a flat holomorphic connection, part of a 
two-dimensional gauge potential. The gauge symmetry is related to
conjugation of the fundamental matrix $\Phi(z)$ by matrices with
rational functions for entries. The gauge invariant quantities
associated to $A(z)$ are the Wilson loop observables
\begin{equation}
W[\gamma_k]={\cal P}\exp\left[\oint_{\gamma_k} A(z)dz\right]=M_{\gamma_k},
\end{equation}
where $\gamma_k$ is a closed loop around the critical point
$z_k$. From the ODE perspective, the matrix $M_{\gamma_k}$ implements
the monodromy around $z_k$
\begin{equation}
\Phi(e^{2\pi i}(z-z_k)+z_k)=\Phi(z)M_{\gamma_k}.
\end{equation}
Because the choice of basis is gauge-dependent, the matrices
$M_{\gamma_k}$ are defined up to conjugation. Their traces are readily
given in terms of the indicial exponents $\alpha_k^\pm$
\begin{equation}
\Tr M_{\gamma_k}=2 e^{\pi
  i(\alpha_k^++\alpha_k^-)}\cos\pi(\alpha_k^+-\alpha_k^-). 
\end{equation}
The sum of the indicial exponents $\alpha_k^++\alpha_k^-$ can be
thought of as an abelian ``charge'': their values do not matter for the
determination of the entries of the monodromy matrix. From the ODE
perspective, their value can be modified with a ``s-homotopic''
transformation, like \eqref{eq:shomos}  and \eqref{eq:shomor}. We will
consider $\alpha_k^++\alpha_k^-=0$ from now  on, and define
$\sigma_k=\alpha_k^+-\alpha_k^-=2\alpha_k^+$.  

Another set of
conjugation-invariant quantities are the composite monodromies
\begin{equation}
p_{kl}=\Tr M_kM_l=2\cos\pi\sigma_{kl},
\end{equation}
which will permit us to solve the problem posed at the end of the last
subsection. Let us define the Frobenius solution $\Phi_k(z)$ the
fundamental matrix such that the first row is given by a pair of
Frobenius solutions constructed at $z=z_k$:
\begin{equation}
\Phi_k(z)=\begin{pmatrix}
y^{+}_k(z) & y^{-}_k(z) \\
w^{+}_k(z) & w^{-}_k(z)
\end{pmatrix},
\end{equation}
where $y^{\pm}_k(z)$ are like \eqref{eq:frobeniussolutions} and the
second row is related to the first by \eqref{eq:wfromy}. It can
then be  verified that the monodromy matrix around $z_k$ is diagonal,
given by entries in \eqref{eq:frobeniusmonodromy}
\begin{equation}
\Phi_k(e^{2\pi i}(z-z_k)+z_k)=\Phi_k(z)M_k,\quad\quad
M_k=\begin{pmatrix}
e^{+\pi i\sigma_k} & 0 \\
0 & e^{-\pi i\sigma_k} 
\end{pmatrix}.
\end{equation}
This is valid for any particular singular point $z_k$ provided we pick
the solution $\Phi_k(z)$ constructed by the corresponding Frobenius
solutions $y_k^\pm(z)$. The hard question to answer is how this matrix
$\Phi_k(z)$  will behave when we take it around a \textit{different}
singular point $z_l$. In order to relate the problem with the
construction of the preceeding paragraphs, let us introduce the
\textit{connection coefficients} between the two Frobenius solutions
\begin{equation}
y^+_k(z)=a_{kl}y^+_l(z)+b_{kl}y^-_l(z),\quad\quad
y^-_k(z)=c_{kl}y^+_l(z)+d_{kl}y^-_l(z).
\end{equation}
Writing these in terms of the fundamental matrix,
$\Phi_k(z)=\Phi_l(z)E_{kl}$, where $E_{kl}$ is the connection
matrix. Now, since the monodromy matrix around $z_l$ is diagonal in
terms of $\Phi_l(z)$, we can perform a change of basis to find
\begin{equation}
\Phi_k(e^{2\pi i}(z-z_l)+z_l)=\Phi_k(z)M_l,\quad\quad
M_l=E_{kl}^{-1}\begin{pmatrix}
e^{+\pi i\sigma_l} & 0 \\
0 & e^{-\pi i\sigma_l} 
\end{pmatrix}E_{kl},
\end{equation}
where $\alpha_l^\pm=\pm \sigma_l/2$ are the two solutions for the indicial equations
at $z=z_l$. Unfortunately, the matrix $E_{kl}$ is in general a very
complicated function of the parameters entering the differential
equation we started with.

For both the problems posed in the last section,
\eqref{eq:boundaryfors} and \eqref{eq:boundaryforr}, we are looking
for parameters of the equations\,---\,\eqref{eq:heunangular} and
\eqref{eq:heunradial}, respectively\,---\,such that there is a special
solution where a given behavior at $z=z_k$ and $z=z_l$ is
specified. This means that one or more connection coefficients\,---\,
entries of the connection matrix $E_{kl}$\,---\,vanish. We will consider
it to be lower triangular then. Let us now choose a basis where the
monodromy matrix at $z=z_k$, $M_k$ is diagonal. Then, by the analysis
above, the monodromy matrix at $z=z_l$, $M_l$ will be the conjugation
of a diagonal matrix by a lower triangular matrix, which will again be
lower triangular. Then it is a direct calculation to verify that the
parameters of the composite monodromy between $z_k$ and $z_l$
introduced above will satisfy the quantization condition
\begin{equation}
\sigma_{kl}=\sigma_k+\sigma_l+2 n,\quad\quad n\in \mathbbold{Z}.
\label{eq:compositemonodromy}
\end{equation}
Another direct calculation shows the converse, if the composite
monodromy parameter $\sigma_{kl}$ satisfies the equation above, then
the monodromy matrices $M_k$ and $M_l$ commute, and hence can be put
simultaneously into the lower triangular form, which in turn means the
vanishing of the connection coefficient and that there are solutions
with the desired prescribed behavior at both points $z_k$ and $z_l$. 

Translating this to the boundary conditions for $S(u)$ and $R(z)$,
\eqref{eq:boundaryfors} and \eqref{eq:boundaryforr}, respectively we
find that, if one can compute the composite monodromy parameter
$\sigma_{kl}$ in terms of the quantities in each ODE
\eqref{eq:heunangular} and \eqref{eq:heunradial}, the normal modes
frequencies would be implicitly given by
\begin{gather}
\sigma_{01}(m_1,m_2,\beta,\Delta,u_0,\lambda_{\ell})=m_1+m_2 +2\ell, 
\quad  \quad \ell \in \mathbbold{Z},\\
\sigma_{+\infty}(\theta_k,\Delta,z_0,
\omega_n,\lambda_\ell)=\theta_\infty+\theta_+ 
+2 n,\quad\quad
n\in \mathbbold{Z}.
\label{eq:implicitsln}
\end{gather}
Some comments are in order:
\begin{itemize}
\item We are thinking of the $\sigma_{ij}$ as a function of the
  parameters in the ODE. In particular, for \eqref{eq:heunangular},
  the list of arguments of the function comprehends
  $m_1,m_2,\beta,q_1,q_2$ from \eqref{eq:angularthetainfty} and the
  accessory parameters $u_0$ and $Q_0$ from
  \eqref{eq:angularaccessory}. Likewise, for \eqref{eq:heunradial},
  these are $\theta_k$ and $\Delta$ from \eqref{eq:radialthetainfty}
  as well as $z_0,K_0$ from \eqref{eq:radialaccessory}. 
\item The first condition pertains to the angular equation, between
  the regular singular points $z=0$ and $z=1$. The second
  to the radial equation between the points $z=r_+^2$ and
  $z=\infty$. In the second one, extra care has to be taken to pick
  the positive (real part of the) frequency solution.
\end{itemize}

In short, the message of the arguments given so far is that the
solution of the problem rests on our hability to calculate the
composite monodromy parameter $\sigma_{ij}$. That is where the
isomonodromy method helps.

The isomonodromy method consists in finding a gauge where $A(z)$ in
\eqref{eq:flatconnection} can be written as a partial fraction expansion
\begin{equation}
A(z)=\frac{A_0}{z}+\frac{A_1}{z-1}+\frac{A_{t}}{z-t}.
\end{equation}
It was demonstrated explicitly in \cite{daCunha:2015uua} that one can
always achieve such a gauge for which the two independent solutions
from first row of the fundamental matrix \eqref{eq:14} satisfy the
Heun equation for generic accessory parameters $t,K_0$. Given the
expression for $A(z)$, we define the $\tau$-function
\begin{equation}
  \frac{d}{dt}\log \tau(t,\vec{\sigma})
  =\frac{1}{t}\Tr(A_0A_{t})+\frac{1}{t-1}  \Tr(A_1A_{t}),
\label{eq:thetaufunction}
\end{equation}
which can be shown \cite{Iwasaki:1991} to be a function of only the
invariant monodromy data like the parameters
$\vec{\sigma}=\{\sigma_k,\sigma_{kl}\}$. For the Heun differential equation
\eqref{eq:heunradial}\,---\,\eqref{eq:heunangular} is entirely analogous
\,---\,the associated function $\tau$ satisfies a version of the Painlevé VI
differential equation, and the composite monodromy data is encoded in
the asymptotics of the solution as $t=0,1,\infty$
\cite{Jimbo:1982}. This differential equation is just the
flat-curvature condition on $A(z)$ stated in terms of each coefficient
matrix $A_i$. These equations are known in the literature as the
Schlesinger equations. 

Now, the $\tau$ function for the Painlevé VI system was shown to
correspond to {\it exact} $c=1$ conformal blocks, and expansions near
$t=0,1$ and $\infty$ were given in
\cite{Gamayun:2012ma,Gamayun:2013auu}, following the proposal
\cite{Alday:2009aq} and subsequent proof \cite{Alba:2010qc} of the AGT
conjecture. Perhaps more tantalizingly, the
same $\tau$-function appears in the {\it semiclassical} expansion for
Liouville conformal blocks \cite{Litvinov:2013sxa}\,---\,see also
\cite{Iorgov:2014vla}. In both cases the role of the Painlevé
$\tau$-function as the character of the Virasoro algebra over a Verma
module is paramount. Given the $\tau$-function expansion, we can
phrase the problem of finding the composite monodromy parameters
implicitly by
\begin{equation}
\begin{aligned}
  \left.
      t(t-1)\frac{d}{dt}\log\tau(t,\vec{\sigma})\right|_{t=z_0}&
  =z_0\theta_{0}\theta_++ (z_{0}-1)\theta_-\theta_{0}  +z_0(z_0-1)K_{0}\\[5pt]
  \left. \frac{d}{dt}\left[t(t-1)
      \frac{d}{dt}\log\tau(t,\vec{\sigma})\right]\right|_{t=z_0}& 
  =(\theta_{-}+\theta_{+}+\kappa_{1})\theta_{0}=
  \frac{\theta_{0}}{2}(\theta_{-}+\theta_{+}-\theta_{0}+\theta_{\infty}), 
\end{aligned}
\label{eq:tauinitialconditions}
\end{equation}
which again comes from the differential equation satisfied by the $\tau$
function, or, in other words, from the flat curvature condition on
$A(z)$ as we change $t$. The conditions
\eqref{eq:tauinitialconditions} apply directly to the radial equation 
\eqref{eq:heunradial}, with a similar set of conditions for the
angular equation \eqref{eq:heunangular}. This deep connection between
massive (scalar) perturbations of Kerr-AdS and Conformal Field Theory
methods is quite surprising and still under development. Exact
expansions for the Painlevé VI $\tau$-function can be found in the
references above and those in the beginning of the subsection. These
expansions, through the conditions \eqref{eq:tauinitialconditions},
achieve an implicit solution to the problem posed in the last Section,
which is to find the parameters $\lambda_\ell$, $\omega_n$, defined in
\eqref{eq:implicitsln}, corresponding to normal modes. We hope to
explore the space of solutions and its implications in future work. 

\section{Semi-classical Conformal Blocks and the near-extremal
  case} 
\label{liouville}

In order to explore the relation between the isomonodromy and
conformal field theory, let us rewrite the radial Heun
equation \eqref{eq:heunradial} in the ${\rm SL}(2,\mathbbold{C})$ form
\begin{equation}
\frac{d^2\Psi}{dz^2}+T(z)\Psi=0,\quad\quad
\Psi(z)=z^{1/2}(z-1)^{1/2}(z-z_0)^{1/2}\Pi(z),
\label{eq:heunslform}
\end{equation}
with $T(z)$ given by
\begin{gather}
T(z)=\frac{\delta_-}{z^2}+\frac{\delta_+}{(z-1)^2}+\frac{\delta_0}{(z-z_0)^2} 
+\frac{\delta_\infty-\delta_--\delta_+-\delta_0}{z(z-1)}
+\frac{z_0(z_0-1)C_0}{z(z-1)(z-z_0)},\\
\delta_k=\frac{1}{4}-\frac{\theta_k^2}{4},\quad\quad k=-,+,0, \quad\quad
\delta_\infty=-\frac{3}{4}-\frac{\mu^2}{4}=\frac{1}{4}-\frac{(\Delta-2)^2}{4},\\
4z_0(z_0-1)C_0=-\frac{\lambda+\mu^2r_0^2-\omega^2}{r_+^2-r_-^2}
-\theta_-^2+\theta_+^2+(2z_0-1)(\theta_0^2-2).
\label{eq:stressenergy}
\end{gather}

We chose the radial equation for the treatment because the concept of
(semi)-classical conformal blocks of Section 2 is particularly useful
here. The form given by \eqref{eq:heunslform} is known in classical
uniformization theory, where the ratio of two linearly independent
solutions $y^{(1)}(z)$, $y^{(2)}(z)$ form the conformal map $w(z)$
between the upper-half plane $z$ and a geodesic quadrangle in the
Poincaré plane, with internal angles
$\phi_i=\pm\pi\sqrt{1-4\delta_i}=\pm \pi\theta_i$. Since our
application has imaginary $\theta_i$ for $z=0,1$, the inner and outer
horizons, the analogy doesn't work there, we will leave the discussion
about these points to the next subsection. In terms of the map
$\zeta(z)=y^{(1)}(z)/y^{(2)}(z)$, equation \eqref{eq:heunslform} is written as
\begin{equation}
\{\zeta(z);z\}\equiv\frac{\partial^3_z\zeta}{\partial_z\zeta}-\frac{3}{2}
\left(\frac{\partial^2_z\zeta}{\partial_z\zeta}\right)^2 =
2T(z),
\label{eq:schwarzianderivative}
\end{equation}
where the left hand side is known as the {\it Schwarzian derivative}.

The uniformization problem can also be cast in terms of classical
field theory. Let us consider again the equation satisfied by any of
the entries of first row of the fundamental matrix $\Phi$, \eqref{eq:14}
\begin{equation}
  \partial_z^2y -(\Tr A + \partial_z\log a_{12})\partial_z y +(\det A
  -  \partial_z a_{11} + a_{11}\partial_z\log a_{12})y =0. 
\end{equation}
We would again like to cast it as a flat-connection condition
\eqref{eq:flatconnection}. Since now the connection is traceless\,---\,
the Wronskian $\det \Phi$ is constant\,---\,then the absence of the first
derivative term requires that $a_{12}$ does not have any zeros on the
complex plane. So we will take it to be a constant
$a_{12}=\mu$. Now, using a Gauss parametrization of $\Phi(z)$ with the
fields $\chi_R,\chi_L,\phi_c$, 
\begin{equation}
\Phi(z)=e^{\chi_L \sigma^-}e^{-\thalf\phi_c \sigma^3}e^{\chi_R\sigma^+}=
\begin{pmatrix}
1 & 0 \\
\chi_L & 1
\end{pmatrix}
\begin{pmatrix}
e^{-\thalf \phi_c} & 0 \\
0 & e^{\thalf\phi_c}
\end{pmatrix}
\begin{pmatrix}
1 & \chi_R \\
0 & 1
\end{pmatrix},
\label{eq:gaussdecomposition}
\end{equation}
we define the holomorphic connection ${\cal
  A}_z=[\partial_z\Phi(z)]\Phi^{-1}(z)$ in the same way as in the
preceding section and its complex conjugate as ${\cal
  A}_{z^*}=({\cal A}_z)^\dagger$. It is a straightforward exercise
to show that the constraint that the upper off-diagonal element is
constant ${\cal A}_{12}=\bar{{\cal A}}_{21}=\mu$\,---\,an
``oper''\,---\,reduces the flat connection condition to the Liouville
equation 
\begin{equation}
\partial_z\partial_{\bar{z}}\phi_c=\mu^2e^{\phi_c},
\end{equation}
encoding the fact that the original metric in $\zeta$ coordinates has
constant negative curvature $-4\mu^2$. The classical solution to the
Liouville equation is given in terms of $\zeta(z)$ 
\begin{equation}
\phi_c(z,{z^*})=\log\left( -\frac{2}{\mu^2}\frac{\partial_z
    \zeta \partial_{z^*}{\zeta^*}}{(\zeta(z)-{\zeta^*}({z^*}))^2}\right).
\label{eq:classicalliouv}
\end{equation}
The field $\phi_c(z,{z^*})$ has obvious singularities when
$\zeta={\zeta^*}$, which is the image of the real line $z={z^*}$. For $z$
on the real line such that $\zeta(z)$ is analytic, the image is composed
of straight lines or circular arcs. However, $\zeta(z)$ has singular
points on the real line, tied to the singular points of
\eqref{eq:heunslform}, where the image develops a wedge or cusp of
angle $\pi\theta_k$\,---\,we will see in the next subsection what happens
for imaginary $\theta_k$. In either case, we can verify that, up to a
translation and a rotation, the map $\zeta(z)$ near $z_k$ behaves as
\begin{equation}
\zeta(z)=(z-z_k)^{\theta_k}(1+\ldots),\quad\quad z\simeq z_k,
\label{eq:asymptow}
\end{equation}
which translates to the asymptotic behavior of the Liouville field via
the classical solution \eqref{eq:classicalliouv}. 

Quantum Liouville field theory is built upon the action 
\begin{equation}
{\cal S}_L=\frac{1}{8\pi
  b^2}\int_{D}d^2\xi[\partial_a\phi_c\partial^a\phi_c+2\mu^2e^{\phi_c}], 
\end{equation}
see \cite{Harlow:2011ny} for a review. The constant $b^2$ has the
interpretation of Planck's constant and the semiclassical limit will
then be defined by $b\rightarrow 0$. The domain of integration $D$
will be taken to be the sphere with flat metric
$\eta_{ab}=\delta_{ab}$. In the usual parametrization of the kinetic
term, we define the Liouville field $\phi=\phi_c/2b$. The theory can
be seen to be conformally invariant at the quantum level if the field
transforms as
\begin{equation}
\phi'(z',(z')^*)=\phi(z,z^*)-\frac{Q}{2}\log\left|\frac{\partial
    z'}{\partial z}\right|^2,
\label{eq:transflaw}
\end{equation} 
where $Q=b+1/b$. The Hilbert space of Liouville field theory
corresponds to highest weight states constructed from the action of
vertex operators $V_{\Delta(\alpha)}=:e^{2\alpha\phi}:$ on a ${\rm
  SL}(2,\mathbb{C})$ invariant ``vacuum'' state $|0\rangle$. The
parameter $\Delta$ is the conformal dimension of the operator and its
dependence on $\alpha$ will be given below. Each of the
$V_{\Delta(\alpha)}$ generates an infinite tower of states by action
of the Virasoro algebra $\{L_n\}$, where
\begin{equation}
T(z)=-(\partial \phi)^2+Q\partial^2\phi=\sum_{n=-\infty}^\infty
L_nz^{-n-2}, 
\end{equation}
satisfying $[L_n,L_m]=(n-m)L_{n+m}+\tfrac{c}{12}\delta_{n+m,0}$, with
$c=1+6Q^2$. Here $T(z)=-2\pi T_{zz}(z)$, the appropriate component of
the stress-energy tensor of the Liouville field, whose conservation
law implies that it is holomorphic. There are analogous formulas for
the anti-holomorphic sector, which we will omit. With some effort
\cite{DiFrancesco:1997}, one can show that $V_{\Delta(\alpha)}$ is a
primary operator, 
\begin{equation}
e^{2\alpha\phi'(z',(z')^*)}=\left(\frac{\partial z'}{\partial
      z}\right)^{-\Delta(\alpha)}\left(\frac{\partial (z')^*}{\partial
      z^*}\right)^{-\bar{\Delta}(\alpha)}e^{2\alpha\phi(z,z^*)},
\end{equation}
where $\Delta(\alpha) =
\alpha(Q-\alpha)=\bar{\Delta}(\alpha)$. Equivalently, one can phrase
the transformation law in terms of the OPE
\begin{equation}
T(z)V_{\Delta}(z_0,z^*_0)\simeq
\frac{\Delta(\alpha)}{(z-z_0)^2}V_{\Delta}(z_0,z^*_0) 
+\frac{1}{z-z_0}\partial_{z_0}V_{\Delta}(z_0,z^*_0). 
\label{eq:tope}
\end{equation}

In the quantum theory, $e^{-\thalf\phi_c}\equiv V_{\Delta(-b/2)}$ corresponds to
a degenerate Virasoro module at level 2, because the operator
\begin{equation}
(L_{-1}^2+b^2L_{-2})e^{-\thalf\phi_c}
\label{eq:nulllevel2}
\end{equation}
has a null norm and therefore must decouple from local operators in
correlation functions.  The corresponding Ward identity 
satisfied by $e^{-\thalf\phi_c}$ is 
\begin{equation}
\langle (L_{-1}^2+b^2L_{-2})e^{-\thalf\phi_c}(z)X(\{z_i\})\rangle=0,
\end{equation}
where $X(\{z_i\})$ is a generic local operator. We will consider the
case where $X(\{z_i\})$ is a product of vertex operators $V_{\Delta_i}$,
and each corresponding $\alpha_i$ of order $1/b$ so it can be thought
of as a ``heavy operator'', setting a background over which
$e^{-\thalf\phi_c}$ reacts. Using the OPE between $T(z)$ and the
primary operator $e^{-\thalf\phi_c}$ from \eqref{eq:tope} one can see
that 
\begin{equation}
L_{-1}e^{-\thalf\phi_c}(z)=\partial_ze^{-\thalf\phi_c}(z),\quad\quad
L_{-2}e^{-\thalf\phi_c}(z)=:T(z)e^{-\thalf\phi_c}(z):.
\end{equation}
Therefore, by considering the subcase where $X$ is the product of four
primary operators, projected to an intermediate channel of conformal
dimension $\Delta$ via a projection operator $\Pi(\Delta)$ (normal
ordering of each operator is implicit), 
\begin{equation}
X(z_a,z_b,z_c,z_d)=
e^{2\alpha_a\phi(z_a)}e^{2\alpha_b\phi(z_b)}\Pi(\Delta)e^{2\alpha_c\phi(z_c)}e^{2\alpha_d\phi(z_d)},  
\end{equation}
and setting $z_a=\infty$, $z_b=z_0$, $z_c=1$ and $z_d=0$, the Ward
identity turns to \eqref{eq:heunslform} in the semiclassical limit
$b\rightarrow 0$. This is a direct consequence of the OPE between
$T(z)$ and the primary operators \eqref{eq:tope}, whose singular terms
will yield \eqref{eq:stressenergy} and the wavefunction
is identified with $\Psi(z)=\langle
e^{-\tfrac{1}{2}\phi_c}X\rangle$. The accessory parameter $C_0$ is given
by the logarithm derivative of the correlation function of $X$
\begin{gather}
C_0=b^2 \frac{d}{d z_0}\log F(z_0),\\
F(z_0)=\lim_{z_a\rightarrow \infty}z_a^{-2\Delta_a}\langle
V_{\Delta_a}(z_a)V_{\Delta_b}(z_0)\Pi(\Delta) V_{\Delta_c}(1)V_{\Delta_d}(0)\rangle
\simeq e^{-S_{\rm cl}(\Delta_i,\Delta)},
\label{eq:accessory}
\end{gather}
where at the last inequality we took the semiclassical approximation
for the $4$-point function as the value of the classical action of the
Liouville field subjected to the boundary conditions at the singular
points \eqref{eq:asymptow}. For the problem of our interest, the role
of the projection operator singles out one intermediate conformal
channel with definite conformal dimension $\Delta$, which we
will take to be ``heavy'': $\Delta = {\cal O}(b^{-2})$, as in
\eqref{eq:semiclassicalblob} below. See Fig. \ref{fig:witten4pt} for a
semiclassical picture. The function $F(z_0)$ is the semiclassical 
conformal block, and could in principle be obtained from the
representation theory of Virasoro algebra. An expansion for small
$z_0$ can be obtained from \cite{Litvinov:2013sxa}. In terms of the
radial equation, a large, and negative, value for $z_0$ is
accomplished near the extremal limit $r_+\rightarrow r_-$:
\begin{equation}
z_0=\frac{r_0^2-r_-^2}{r_+^2-r_-^2}
=-\frac{r_+^2+2r_-^2+a^2+b^2+1}{r_+^2-r_-^2} 
\end{equation}
By the same token, a large intermediate dimension means a large
intermediate monodromy parameter, which in turn can be heuristically
assigned to a finite frequency in \eqref{eq:stressenergy}. To sum up,
the semi-classical approximation is then expected to describe the
radial equation in the near-extremal, finite frequency limit. 

As an important comment, in the semiclassical limit considered below,
the relation between the accessory parameter $C_0$ and the conformal
dimension of the intermediate channel is a bijection, as expressed in
relations obtained in the preceding Section like
\eqref{eq:implicitsln}. From this one deduces that the relevant
quantity for finding the normal modes of \eqref{eq:heunslform} is not
exactly the generic four-point function but the conformal block. The
integer factors arising in the expansion correspond to the Virasoro
descendants of the intermediate channel.

\subsection{Entropy and Liouville momenta}

The classical asymptotics of the Liouville field
\eqref{eq:asymptow} means that the ``physical'' metric
\begin{equation}
ds^2=e^{\phi_c}dz\,dz^*
\end{equation}
will develop a cusp-like singularity at $z=z_k$ for real
$\theta_k$. This is indeed the case for the singularities at
$z=\infty$ and $z=z_0$ in the radial equation
\eqref{eq:heunradial}. The latter is real due to the fact that $r_0$
is purely imaginary and therefore so is the ``temperature'' $T_0$. For
the angular equation \eqref{eq:heunangular}, all the singularities
have real indicial exponents, so the semiclassical interpretation of
the corresponding four-point function as living on a sphere is still
valid. As discussed above, the geometric interpretation in these 
cases is of a deficit angle $\pi\theta_k$ at $w_k=w(z_k)$, and the
singular points are related to local operators
$:e^{2\alpha_k\phi(z_k)}:$ with scaling dimension 
\begin{equation}
\delta_k=b^2\Delta_k=\eta_k(1+b^2-\eta_k)\simeq
\eta_k(1-\eta_k), 
\label{eq:semiclassicalblob}
\end{equation}
where $\eta_k=b\alpha_k$.  For $\eta_k$ real and smaller than $1/2$,
the geometry of the physical metric develops a cusp at $z_k$. These
will be associated with the singular points at $z=\infty$ and
$z=z_0$. From the relation between the Liouville field and the map
$\zeta(z)$, we learn that the deficit angle $\pi\theta_k=2\eta_k$, and
hence vertex operators with $\alpha_k>1/2b\simeq Q/2$ do not have a
semiclassical geometrical interpretation. This is called the {\it
  Seiberg bound}. At the quantum mechanical level, we learn\,---\,see,
for instance, \cite{Harlow:2011ny}\,---\,that
\begin{equation}
V_{\Delta(\alpha)}=C(\alpha)V_{\Delta(Q-\alpha)},
\label{eq:reflectionproperty}
\end{equation}
where $C(\alpha)$ is called the ``reflection amplitude'' and is a
known function of $\alpha$. 

For the radial equation \eqref{eq:heunradial}, the singularities at
$z=z_0$ and $z=\infty$ have real $\delta_k$ associated to them. The
picture above is well suited provided we use the reflection property
\eqref{eq:reflectionproperty} to assign the singularity to an
operator\,---\,possibly heavy\,---\,with $\alpha\approx Q$. For
instance, the insertion at 
$z=\infty$ can be seen to correspond to a ``heavy operator''
\cite{Hijano:2015zsa,Fitzpatrick:2015zha}
$\alpha_\infty=\eta_\infty/b$ with  
\begin{equation}
\eta_\infty = \frac{\Delta-1}{2},\frac{3-\Delta}{2},
\end{equation}
with the first solution being proper for $\Delta<2$ and the second for
$\Delta>2$, see \cite{Harlow:2011ny} for a detailed discussion.

On the other hand, the singularities at
$z=0,1$, corresponding to the inner and outer horizon have the
property that 
\begin{equation}
\theta_\pm =\frac{i}{2\pi}\left(\frac{\omega -m_1\Omega_{\pm,1}-m_2
    \Omega_{\pm,2}}{ T_\pm}\right) = \frac{i}{2\pi}\delta S_\pm, 
\end{equation}
where $\delta S_\pm$ is the variation of entropy of the black hole
horizon at $r=r_\pm$, respectively, due to the admittance of energy
$\omega$ and angular momenta $m_1$ and $m_2$. $\delta S_\pm$ is then
real and finite for physically reasonable parameters black hole
parameters $M$, $a_1$ and $a_2$. 

\begin{figure}[htb]
\begin{center}
\includegraphics[width=0.6\textwidth]{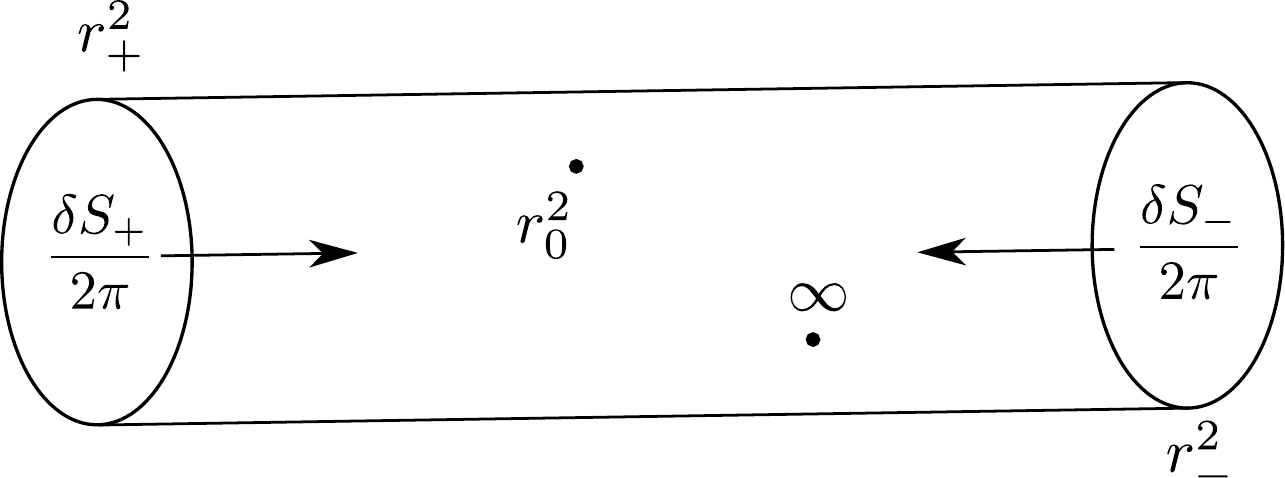}
\caption{The ``world-sheet'' Liouville description of the conformal
  block related to the black-hole scattering. While the insertions of
  the angular equation \eqref{eq:heunangular} are all cusp-like
  singularities, the radial equation \eqref{eq:heunradial} possesses
  conformal boundaries related to the inner and outer horizon. The
  effective geometry is that of a cylinder, and the entropy increase
  of the horizons is equal to the Liouville momentum inserted at each
  horizon.}
\label{fig:liouville4pt}
\end{center}
\end{figure}

In this case the ``physical'' metric near $z_\pm=0,1$ is approximately
given by
\begin{equation}
ds^2=e^{\phi_c}dz\,dz^*=\frac{1}{2\mu^2}
\frac{dz'\,d(z')^*}{\sin^2(\frac{\delta S_\pm}{2\pi}\frac{z'+(z')^*}{2})},
\end{equation}
where we made the transformation $z-z_\pm=e^{z'}$ in the
second step ($z_\pm=0,1$). The approximation is valid for $z'$ sufficiently
close to zero. Because of the identification of the imaginary part of
$z'$, the metric is actually defined on a cylinder. At the quantum
level, the vertex operator at $z_\pm=0,1$ corresponds to
$V_{\Delta(\alpha_\pm)}(z_\pm)$, with 
\begin{equation}
\alpha_\pm=\frac{Q}{2}+i\frac{\delta S_\pm}{4\pi b},
\end{equation}
and, in our conventions, we consider $z$ to be the complex plane. If
we map the plane to the cylinder by $w=\log z$, the Liouville momentum
\,---\,the zero mode for $\Pi=\dot{\phi}$\,---\,will be shifted
because of the anomalous transformation law \eqref{eq:transflaw}
\begin{equation}
\partial_z\phi\rightarrow \frac{1}{z}(\partial_w\phi-Q/2),
\end{equation} 
and then $\delta S_\pm/2\pi$ is the Liouville momentum injected at
$z_\pm$. We illustrate the Liouville diagram computed by the conformal
block in Fig. \ref{fig:liouville4pt}.

The geometrical picture actually helps us writing an approximate
expansion for the composite monodromy in the near horizon limit
$r_+\rightarrow r_-$. This
limit is characterized not only by a large\,---\,and negative value of
$z_0$, but also large values for $\theta_+$ and $\theta_-$, which
scale as $z_0$. Although there are no ``light'' operators, there is a
hierarchy because $\theta_0$ and $\delta_\infty$ do not
scale with $z_0$. We can combine a method given by
\cite{Zamolodchikov:1986aa} and notions of thermality to write an
approximate expression for the accessory parameter $C_0$ in terms of
the composite monodromy parameter between $z=1$ and $z=\infty$,
which we will call $\delta_{1,\infty}=\delta$.  The limit is
complicated by the generic scaling behavior:
\begin{equation}
\delta_\pm = {\cal O}(z_0^2),\quad
\delta_+-\delta_- = {\cal O}(z_0),\quad
\delta_0,\Delta = {\cal O}(z_0^{0}),\quad
C_0 = {\cal O}(z_0^{-1}).
\end{equation}
By transforming $u=1/z$, we bring the singular points of interest to
$0$ and $1$. The equation \eqref{eq:heunslform} is
\begin{equation}
\frac{d^2\bar{\Psi}}{du^2}+\bar{T}(u)\bar{\Psi}=0,\quad\quad
\bar{\Psi}(u)=u\Psi(z=1/u),
\label{eq:heuninverted}
\end{equation}
with 
\begin{gather}
\bar{T}(u)=\frac{\delta_\infty}{u^2}+\frac{\delta_+}{(u-1)^2}+
\frac{\delta_0}{(u-u_0)^2} 
+\frac{\delta_--\delta_\infty-\delta_+-\delta_0}{u(u-1)}
+\frac{u_0(u_0-1)\tilde{C}_0}{u(u-1)(u-u_0)},\\
u_0=\frac{1}{z_0},\quad\quad
\tilde{C}_0=-z_0^2C_0-2z_0\delta_0.
\end{gather}

Now we observe that the solution $\bar{\Psi}(u)$ of
\eqref{eq:heuninverted} corresponds to the classical
profile of the Liouville field $e^{-\thalf \phi_c}$ in the presence of
the ``heavy operators'' $V_{\Delta_i}$. Upon a conformal
transformation $\tilde{u}(u)$, it is straightforward to verify that
$\tilde{\Psi}(\tilde{u})=(\partial_u\tilde{u})^{1/2}\bar{\Psi}(u)$ satisfies:
\begin{equation}
\partial_{\tilde{u}}\tilde{\Psi}+\tilde{T}(\tilde{u})\tilde{\Psi}=0
\end{equation}
where
\begin{equation}
\tilde{T}(\tilde{u})=(\partial_u\tilde{u})^{-2}\left(\bar{T}(u)-\frac{1}{2}\{\tilde{u};u\}\right).
\end{equation}

By choosing $\tilde{u}$ such that:
\begin{equation}
\frac{1}{2}\{\tilde{u};u\}=\frac{\delta_\infty}{u^2}+\frac{\delta_+}{(u-1)^2}+
\frac{\delta_--\delta_\infty-\delta_+-\delta_0}{u(u-1)}
\label{eq:triangleeq}
\end{equation}
one ``reduces'' $\tilde{T}(\tilde{u})$ to two terms:
\begin{equation}
\tilde{T}(\tilde{u})=(\partial_u\tilde{u})^{-2}\left(\frac{\delta_0}{(u-u_0)^2}+
\frac{u_0(u_0-1)\tilde{C}_0}{u(u-1)(u-u_0)}\right),
\end{equation} 
and, since the second term is large in the large $z_0$ limit, we can
give an approximate solution for $\tilde{\Psi}$ using a WKB
approximation:
\begin{align}
\tilde{\Psi}(\tilde{u}) & =\tilde{\Psi}_0\exp\left[\pm\int^{\tilde{u}}_{\tilde{u}_i}d\tilde{u}'
(\partial_u\tilde{u}')^{-1}
\sqrt{\frac{\delta_0}{(u(\tilde{u}')-u_0)^2}+
\frac{u_0(u_0-1)\tilde{C}_0}{u(\tilde{u}')(u(\tilde{u}')-1)(u(\tilde{u}')-u_0)}}
 \right] \\
& = \tilde{\Psi}_0\exp\left[\pm\sqrt{\delta_0}\int^{u}_{u_i}\frac{du'}{u-u_0}
\sqrt{\frac{(u'-u_+)(u'-u_-)}{u'(u'-1)}}
 \right],
\label{eq:psitilde}
\end{align}
where we assumed $\partial_u\tilde{u}\neq 0$ in the path of
integration. The constants $u_\pm$ are given by:
\begin{equation}
u_\pm = -\frac{D+1-u_0(D+2)}{2}\left[
1\pm\sqrt{1+\frac{4u_0(1-u_0)(D+2)}{(D+1-u_0(D+2))^2}}\right],
\quad D=\frac{C_0}{u_0\delta_0}.
\end{equation}
One notes that $D$ is of order ${\cal O}(z_0^0)$. For small $u_0$,
one has:
\begin{equation}
u_+=-(D_0+1)+{\cal O}(u_0),\quad
u_-=\frac{D_0+2}{D_0+1}u_0 +{\cal O}(u_0^2),
\end{equation}
where $D_0=\lim_{u_0\rightarrow 0}C_0/u_0\delta_0$. Both $u_\pm$ are
negative for large enough frequency.  

Now, equation \eqref{eq:triangleeq} can be solved explicitly
in terms of hypergeometric functions:
\begin{equation}
\tilde{u}=
\frac{u^{\tfrac{1}{2}(\theta_+-h)}(u-1)^{\tfrac{1}{2}(1-\theta_+)}
{_2F_1}(\tfrac{1}{2}(1+\theta_\infty-\theta_++h),\tfrac{1}{2}(1-\theta_\infty-\theta_++h);
1-h;u^{-1})}{u^{\tfrac{1}{2}(\theta_++h)}(u-1)^{\tfrac{1}{2}(1-\theta_+)}
{_2F_1}(\tfrac{1}{2}(1+\theta_\infty-\theta_+-h),\tfrac{1}{2}(1-\theta_\infty-\theta_+-h);
1+h;u^{-1})}
\end{equation}
where $h^2=1+\theta_-^2-\theta_0^2$ and $\theta_\infty=\Delta-2$. The
function $\tilde{u}$ is defined up to a Möbius map, stemming from the
choice of hypergeometric functions in the numerator and
denominator. The choice above guarantees that we have positive
Liouville momentum entering $u=u_0$ and that we have a diagonal
monodromy near $u=\infty$. The function $\tilde{u}$ contributes the
term:
\begin{equation}
(\partial_u\tilde{u})^{-1/2}=W^{-1/2}u^{\tfrac{1}{2}(\theta_++h)}(u-1)^{\tfrac{1}{2}(1-\theta_+)}
{_2F_1}(\tfrac{1}{2}(1+\theta_\infty-\theta_+-h),\tfrac{1}{2}(1-\theta_\infty-\theta_+-h);
1-h;u^{-1}),
\label{eq:jacobianhyperg}
\end{equation}
to the solution $\bar{\Psi}(u)$, where $W$ is a constant. 

From the splitting
$\bar{\Psi}(u)=(\partial_u\tilde{u})^{-1/2}\tilde{\Psi}(\tilde{u}(u))$ we
have an approximate solution for \eqref{eq:heunslform}. Calculating
the monodromy between the points $u=1$ and $u=0$, corresponding
respectively to the outer horizon and infinity is then a
straightforward exercise in complex analysis
\cite{Zamolodchikov:1986aa}. The integral in \eqref{eq:psitilde} is
related to incomplete elliptic integrals, which for the contour $C_{01}$
around $u=0$ and $u=1$ picks a complete elliptic integral of the
third kind $\Pi(\nu,k)$:
\begin{multline}
\oint_{C_{01}}\frac{du}{u-u_0}\sqrt{\frac{(u-u_+)(u-u_-)}{u(1-u)}}
= -\frac{4u_-(u_+-u_0)}{u_0\sqrt{u_+(u_--1)}}
\Pi\left(\frac{u_--u_0}{u_0(u_--1)},\sqrt{\frac{u_+-u_-}{u_+(1-u_-)}}\right) 
\\
-\frac{4u_-}{\sqrt{u_+(u_--1)}} \Pi\left(
\frac{1}{1-u_-},\sqrt{\frac{u_+-u_-}{u_+(1-u_-)}}\right).
\end{multline}
For small $u_0$, we use the expansion:
\begin{equation}
\Pi(\nu,k^2)=\frac{1}{1-\nu}\left[\left(1+{\cal
      O}(k'^2)\right) \log\frac{4}{k'^2}
-\sqrt{\nu}\log\frac{1+\sqrt{\nu}}{\sqrt{|1-\nu|}}
\right]
\end{equation}
with $k'^2=1-k^2$ and $0<\nu<1$, and we have the asymptotic behavior:  
\begin{equation}
\oint_{C_{01}}\frac{du}{u-u_0}\sqrt{\frac{(u-u_+)(u-u_-)}{u(1-u)}}=
-2\frac{D_0+2}{D_0}\sqrt{|D_0+1|}\log(-u_0)+{\cal O}(u_0, u_0\log(-u_0))
\end{equation}
At this point one has to check whether the WKB approximation is valid
on the contour $C_{01}$. The conditions for the validity were considered by
\cite{Winitzki:2005rw}, where it is shown that the precision of the 
WKB series is tied to the existence of a zero or singularity of the
integrand at some $\bar{u}$ where the divergence is milder than
$|u-\bar{u}|^{-1+\epsilon}$, with $\epsilon>0$.  In our application
this is not the case near $u_0$, but it is near $u_+,u_-,0,1$. For the
contour $C_{01}$, the deformation should be valid for intermediate
ranges of $u_0$ or in those cases where $u_->u_0$ -- which requires
$D_0<-1$. In any case, the contour can be deformed in the complex
plane to avoid the pole at $u_0$ and errors can be estimated to be of
order $u_0\log(-u_0)$.

Finally, one notes that, while the solutions
$\bar{\Psi}(u)$ in \eqref{eq:heuninverted} is analytic everywhere
except at the critical points $u=0,u_0,1,\infty$, the approximation
$\tilde{\Psi}(\tilde{u}(u))$ has extra branching points at
$u=u_\pm$. Thus, the approximation used here displays the {\it Stokes
  phenomenon}: each of the solutions at \eqref{eq:psitilde} can only
approximate the exact solution on regions radiating from
$u_\pm$. Since $u_\pm$ is a simple zero in the radical of the
integrand, the situation here is completely analogous to the Stokes
phenomenon arising from the WKB approximation of the Airy function
near a classical turning point:
the principal region is limited to the anti-Stokes line $|\arg
(u-u_\pm)|<\pi/3$. It can be seen that the contour encompassing
$u=0,1$ considered can be seen to keep to the principal
region, since we can deform $C_{01}$ to cross the real line in between
$u_-$ and $0$. In the classical mechanics analogy, we are never going
over the classical turning point by following through
$C_{01}$. Therefore each of the two solutions \eqref{eq:psitilde} 
are good throughout the contour, and no extra Stokes parameters are
picked. 

With these provisions, the contribution for the monodromy from the
$(\partial_u\tilde{u})^{-1/2}$ term is readily computed. Since it is given by a
hypergeometric function, we know that a curve encompassing $u=0$ and
$u=1$ can be deformed to a curve going around $u=\infty$, and the
monodromy phase then equals $i\pi(1\pm h)$ from the exponent of
the $u$ term in \eqref{eq:jacobianhyperg}. 

The CFT interpretation of this composite monodromy stems from the OPE
between the ``light'' operator
$e^{\tfrac{1}{2}\phi_c}=V_{\Delta(-b/2)}$ and the intermediate
conformal channel obtained from the fusion between two of the
operators in \eqref{eq:accessory}:
\begin{equation}
V_{\Delta_a}(u_a)V_{\Delta_b}(u_b)=(u_a-u_b)^{\Delta-\Delta_a-\Delta_b}
\left( V_{\Delta}(u_a)+{\cal O}((u_a-u_b))\right).
\end{equation}
Now, since the light operator $V_{\Delta(-b/2)}$ is associated to a
degenerate representation\,---\,it has a null descendant of level 2
\eqref{eq:nulllevel2}\,---\,then the OPE between it and $V_{\Delta}$
  has only two channels: $\Delta_\pm=\alpha_\pm(Q-\alpha_\pm)$, where
  $\alpha_\pm=\alpha \pm b/2$. Near $u_a$ we have again:
\begin{equation}
V_{\Delta}(u_a)V_{\Delta(-b/2)}(u)=(u-u_a)^{\Delta_\pm-\Delta-\Delta(b/2)}V_{\Delta_\pm}+\ldots
\end{equation}
and, assuming that $V_{\Delta}(u_a)$ is heavy, we define
$\delta=b^2\Delta$ and in the semiclassical limit the behavior near
$u_a$ is approximately given by
$u^{\tfrac{1}{2}(1\pm\sqrt{1-4\delta})}$. Defining $\sigma$ as
$4\delta=1-\sigma^2$, we can write the monodromy
picked around $u_a\rightarrow 0$ as the phase $i\pi(1\pm\sigma)$. 

Equating both results, we arrive at the result for the composite
monodromy parameter $\sigma$ for large negative $z_0$ and $D_0<-1$ as
defined above
\begin{equation}
\sigma = h+\frac{2}{i\pi}\frac{z_0C_0+2\delta_0}{z_0C_0}
\sqrt{|z_0C_0+\delta_0|}
\log(-z_0)+{\cal
  O}((-z_0)^{-1},(-z_0)^{-1}\log(-z_0)),
\end{equation}
where the signs were chosen to recover the right limit as
$C_0$ vanishes. For large $C_0$, the $\delta_0$ factor can be
disregarded. As we can see from \eqref{eq:stressenergy}, this
result is consistent with the fact that the intermediate channel
dimension is large for large $z_0$.  This result can be used to
determine the spectrum of quasi-normal modes in the limit where
$\delta S_\pm$ is large. We will postpone this study to future work.

\section{Discussion}

In this paper we reviewed the relation between four dimensional
classical conformal blocks and the propagation of fields in AdS
spaces. We also estabilished the relation of the scattering problem of
a scalar field in a Kerr-AdS background in five dimensions to two
dimensional conformal blocks, and argued that the near-extremal limit
of the black hole is described by semiclassical Liouville conformal
blocks, in a picture somewhat different from \cite{Siopsis:2004as}. We
also saw that singular points of the radial equation are physically tied to
the positions of the horizons, the point at infinity and the negative
root of $\Delta_r$ \eqref{eq2}. While the points are related to usual
vertex operators in Liouville field theory, we argued that the
singular points at the inner and outer horizon saturate the Seiberg
bound, with the Liouville momentum of the operator given the variation
of the entropy of the black hole horizon by absorption of the incoming
wave. This gives a two-dimensional geometrical description of the
process, as the Liouville correlator on the cylinder with a number of
marked points\,---\,see Fig. \ref{fig:liouville4pt}. Remarkably, the
corresponding Liouville description seems unitary in itself. 

It would be very interesting to see whether this classical
correspondence also holds at the quantum level. Features like the
central charge have been shown several times to be related to the
black hole entropy, so it is tempting to see whether one can push the
analogy further and recover information from the quantum theory from
the description provided here. Also, the intringuing way
two-dimensional CFTs keep appearing in higher dimensional problems may
herald some deeper connection to quantum gravity. Interestingly, the
same connection between CFT and integrability is also frequent, common
and mysterious. Mathematically, it is a bit of a surprise that the
full Virasoro algebra\,---\,with $c=1$\,---\,seems relevant to the
connection problem of the four-singularity Fuchsian equation. All of
these points deserve better understanding.

Lastly, the analytical tools developed here, specially the inspection
of \eqref{eq:implicitsln}, may help understand the constraints posed
by holography to the RG flow in a generic setting. As important
properties of non-abelian gauge theories, such as confinement and the
conformal window, depend crucially on the specific way the theory
flows from the UV to the IR, it is important to understand in detail
which flows can be holographically described. Implementation of more
realistic elements such that spinorial fields and flavor should be
still mathematically tractable by the tools developed
here. Mathematically the spinor equation on a Kerr-AdS background has
an entirely analogous treatment in terms of Heun's equations as
described. Also, with controllable parameters even at $T=0$\,---\,the
extremal case, one can study the zero temperature limit. Since in the
zero temperature a conformal group is still present 
\cite{Guica:2008mu}, there should be still a massless sector
there. The transition may further help shed light on the physical properties
listed above from a holographic perspective, taking into account
recent progress on QCD and SUSY QCD made in
\cite{Bochicchio:2017sgq,Bochicchio:2016euo}. 

\section*{Acknowledgements}

The authors are greatly thankful to M. Guica for discussions,
ideas and suggestions, and for the referee for spotting an error in
the previous calculation for the composite monodromy in the near
extremal limit. We would also like to thank F. Novaes,
A. Queiroz, F. Rudrigues, D. Crowdy, G. Vasconcelos. BCdC is thankful
to PROPESQ/UFPE and FACEPE for support under grant
no. APQ-0051-1.05/15. 


\providecommand{\href}[2]{#2}\begingroup\raggedright\endgroup

\end{document}